\def\rfr#1{eq. (\ref{#1})}

\def\asec{$''$ cy$^{-1}$}

\def\dert#1#2{\frac{{{d}}{#1}}{{{d}}{#2}}}              

\def\asec{$''$ cy$^{-1}$}

\def\bar{\begin{eqnarray}}
\def\ear{\end{eqnarray}}

\def\eqi{\begin{equation}}
\def\eqf{\end{equation}}
\def\eqia{\begin{eqnarray}}
\def\eqfa{\end{eqnarray}}
\def\rp#1#2{{#1\over#2}}

\def\lb#1{\label{#1}}

\def\oc2{$\mathcal{O}(c^{-2})$}
\def\yu{\exp\left(-\rp{r}{\lambda}\right)}

\documentclass[11pt]{article}
\usepackage{amsmath,amsthm,amscd,amssymb}
\usepackage{latexsym}
\usepackage{graphicx,epsfig}
\usepackage{natbib}
\usepackage{ifpdf}
\ifpdf
\usepackage[%
  pdftitle={Instructions for use of the document class
    elsart},%
  pdfauthor={Simon Pepping},%
  pdfsubject={The preprint document class elsart},%
  pdfkeywords={instructions for use, elsart, document class},%
  pdfstartview=FitH,%
  bookmarks=true,%
  bookmarksopen=true,%
  breaklinks=true,%
  colorlinks=true,%
  linkcolor=red,anchorcolor=red,%
  citecolor=blue,filecolor=red,%
  menucolor=red,pagecolor=red,%
  urlcolor=cyan]{hyperref}
\else
\usepackage[%
  breaklinks=true,%
  colorlinks=true,%
  linkcolor=red,anchorcolor=red,%
  citecolor=blue,filecolor=red,%
  menucolor=red,pagecolor=red,%
  urlcolor=cyan]{hyperref}
\fi

\begin{document}

\noindent{\bf \LARGE{Constraints on the  range $\lambda$ of
Yukawa-like modifications to the Newtonian inverse-square law of
gravitation from Solar System planetary motions}}
\\
\\
\\
{Lorenzo Iorio}\\
{\it Viale Unit$\grave{a}$ di Italia 68, 70125\\Bari(BA), Italy
\\tel. 0039 328 6128815
\\e-mail: lorenzo.iorio@libero.it}

\begin{abstract}
In this paper we use the latest corrections to the Newton-Einstein secular  rates of perihelia of some planets of the Solar
System, phenomenologically estimated with the EPM2004 ephemerides by the Russian astronomer E.V. Pitjeva, to put severe constraints on the range parameter $\lambda$
characterizing the Yukawa-like modifications of the Newtonian
inverse-square law of gravitation. It turns out that the range
 cannot exceed about one tenth of an Astronomical Unit.
 We assumed neither equivalence principle violating effects nor spatial variations of $\alpha$ and $\lambda$. This finding may have
important consequences on all the modified theories of gravity
involving Yukawa-type terms with range parameters much larger than
the Solar System size. However, caution is advised since we, currently have at our disposal only the  extra-rates of periehlia estimated by Pitjeva: if and when other groups will estimate their own corrections to the secular motion of perihelia, more robust and firm tests may be conducted.
\end{abstract}



\section{Introduction}
Historically, the first attempts to find deviations from the Newtonian inverse-square  law of gravitation were performed to explain the anomalous secular
precession of Mercury's perihelion discovered by \citet{Lev62}: \citet{Hal1894} noted that he could account for Mercury's precession if the law of gravity, instead of falling off as $1/r^2$, actually falls of as $1/r^k$ with $k=2.00000016$. However, such an idea was not found to be very appealing, since it conflicts with basic conservation laws, e.g., Gauss's Law, unless one also postulates a correspondingly modified metric for space.   Other historical attempts to modify Newton's law of gravitation to account for the Mercury's perihelion behavior yielded
velocity-dependent additional terms: for a review of them see \citep{Gin07} and references therein.  Such attempts practically ceased after the successful explanation of the perihelion rate of Mercury by \citet{Ein15} in terms of his general theory of relativity: an exception is represented by \citet{Man30} who, with a $1/r^2$ correction to the Newtonian potential, was able to reproduce the anomalous apsidal precession of Mercury.

It was recently realized that deviations from the Newton's inverse-square law could provide windows into new physics \citep{Fuj91,Fis92}.
Indeed, in the modern framework of the challenge of unifying gravity with the other three fundamental forces of Nature
possible new phenomena could show up as deviations from the inverse-square law of gravitation. In general, they would occur
at submillimeter length scales, but sometimes also at astronomical or even cosmological distances.  For a review of the many theoretical speculations
about deviations from the $1/r^2$ law see \citep{Ade03}.

Among various parameterizations like, e.g., power-law \citep{Fis01}, a very popular, phenomenological way to account for a possible
violation of the Newtonian inverse-square law  takes the form of a Yukawa-like, exponentially modified
Newtonian potential \eqi U=-\rp{GM}{r}\left[1+\alpha \exp\left(
-\rp{r}{\lambda}\right)\right],\lb{ypot}\eqf where $G$ is the
Newtonian gravitational constant, $M$ is the mass of the central
body which acts as source of the gravitational field, $\alpha$ and
$\lambda$ are the strength\footnote{Here we will not consider
composition-dependent $\alpha$ which would induce violations of
the equivalence principle; for a derivation of the potential of \rfr{ypot} in a relativistic gravity model obeying the equivalence principle see \citep{Zhy94}.} and the range, respectively, of the
putative new interaction. The Yukawa correction to the Newtonian potential
\eqi U_{\rm
Y}=-\rp{GM\alpha}{r}\exp\left(-\rp{r}{\lambda}\right)\eqf yields
an entirely radial extra-acceleration \eqi A_{\rm
Y}=-\rp{GM\alpha}{r^2}\left(1+\rp{r}{\lambda}\right)\yu.\lb{yacc}\eqf
For a review of various theoretical frameworks
(braneworld models, scalar-tensor or scalar-tensor-vector theories of gravity, studies of topological defects) yielding a Yukawa-like,
fifth force  see, e.g., \citep{Kra01,Ber05,Ber06,Mof06a} and references
therein. Among them, there are various models of modified gravity which predict effects at  astronomical scales or even larger.
For example, the recent Scalar-Tensor-Vector Gravity (STVG) by \citet{Mof06a}
in the intentions of his proponent would be able to
comprehensively and consistently account for the observed data in
the Solar System, the Galaxy, clusters of galaxies and
cosmological scenarios \citep{Mof06b}.
Other studies on long-range, Yukawa-like modifications of gravity  conducted with different techniques on astronomical/astrophysical scales can be found in \citep{Whi01,Ame04,Sea05,Rey05,Shi05,Ser06a}.

The problem of finding experimental or observational constraints
on the parameters $\alpha$ and $\lambda$, which is
crucial  to exclude unviable models and achieve some progress in the study on those that
appear feasible, is usually tackled by
looking at what happens at $\alpha$ by keeping $\lambda$ fixed,
and subsequently repeating the process by sampling different
spatial ranges for $\lambda$, without
asking if this or that particular range for $\lambda$ is, in fact,
really allowed: see, e.g. \citep{Mik77,DeR86,Bur88,Fis99,Kra01,Ber03,Ser06b,Ior07a}.
%

In this paper we will test, in a purely phenomenological way, a
very definite and widely used assumption in many modified theories
of gravity, i.e. the hypothesis that $\lambda$ may assume values
of the same order of magnitude, or larger than the typical sizes
of the planetary orbits in the Solar System \citep{Mof06b}.
We will show that
%
%
Solar System tests are, in fact, able to
tell us something important about ranges $\lambda \gg 10^{11}$ m. To this aim, we will,
first, derive an explicit expression of the secular, i.e. averaged
over one orbital revolution, perihelion precession induced by a
Yukawa-type anomalous acceleration on the orbits of the Solar
System planets. Then, we will compare
our formula, obtained in the small eccentricity approximation,
with the latest estimated corrections of the perihelion\footnote{The perihelia, as the other Keplerian orbital elements, are not observable quantities: ranges, range-rates, right ascensions, declinations are, in fact, measured.} rates \citep{Pit05a} in order to see if results
consistent with the tested hypothesis are obtained. About the methodology adopted, it is important to note that the corrections to the perihelion rates determined in \citep{Pit05a} are phenomenologically estimated quantities of a global, least-square solution in which only Newtonian and Einsteinian dynamics was modeled: no exotic dynamical terms were included in the fit. Thus, in our opinion, such phenomenological corrections can genuinely be used to get information on a hypothetic, unmodeled force. It may be interesting to note that a somewhat independent test of the reliability of such a strategy can be found in \citep{Ior07b} in which the mass of the Kuiper Belt Objects was assessed with the extra-rates of perihelia by \citet{Pit05a} by obtaining results compatible with other estimates from different, non-dynamical techniques. If and when other groups will estimate their own corrections to perihelion rates we will use such determinations as well in order to enforce and extend our test.
A complementary approach which could be followed consists in repeating the global fit of the Solar System data by modifying the dynamical force models of the data reduction softwares with the addition of the investigated non-standard acceleration term and, accordingly, including in the set of the parameters to be estimated in the least-square sense the ones connected with the Yukawa potential as well, so to look at their mutual correlations as well\footnote{According to some people, this would be the only trustable approach to the problem.}:  however, such a strategy would be model-dependent and might yield just the outcomes desired by the experimentalist.

\section{The effects of a Yukawa-like fifth force on the perihelia}
To be more definite, let us suppose that a
given theory, for various theoretical and/or observational
reasons, makes use of a $\lambda$ quite larger than the typical
spatial scales of the Solar System, e.g. because of a fit of a
data set of a physical system different from it. In this case, an
independent test of such an assumption is to check if a $\lambda$
with such characteristics yield, in fact, results compatible with the determined Solar System dynamics, within the associated errors.
Clearly, should un-physical and/or
inconsistent results be obtained, the considered model(s) and the
related hypothesis would be ruled out.

Let us, now, work out the orbital effects induced by \rfr{yacc},
treated as a small perturbation of the Newtonian monopole term, on
the planetary motions of the Solar System planets. In view of a
direct comparison with the latest estimated extra-rates of the longitude of perihelion
$\varpi$, we
will consider the secular precession of such an element. For a radial perturbing acceleration $A_r$, the Gauss
equation for the variation of $\varpi$ can be written as
\eqi\dert\varpi t=-\rp{\sqrt{1-e^2}}{nae}A_r\cos f,\lb{gauss}\eqf
where $a$ is the planet's semimajor axis, $e$ is the eccentricity,
$n=\sqrt{GM/a^3}$ is the Keplerian mean motion and $f$ is the true
anomaly. In order to obtain the secular rate of $\varpi$,
\rfr{yacc} must, first, be evaluated upon the unperturbed Keplerian
ellipse, given by \eqi r=a(1-e\cos E) \eqf in terms of the
eccentric anomaly $E$; then, it must be inserted into the
right-hand-side of \rfr{gauss} and, finally, the integral over a
complete orbital revolution must be performed. The
following formulas will be used
\begin{equation}\left\{\begin{array}{lll}
\cos f=\rp{\cos E-e}{1-e\cos E},\\\\
dt=\rp{(1-e\cos E)}{n}dE,
\lb{FORMULAS}\end{array}\right.\end{equation} In the calculation,
which we  are going to perform by quite reasonably assuming that
$\alpha$ and $\lambda$ are constant and uniform over the typical
spatial and temporal scales of Solar System bodies, the expression
\eqi\exp\left(\rp{ae}{\lambda}\cos E\right)\lb{espo}\eqf appears;
it prevents us from obtaining a closed form of the averaged
perihelion rate because the modified Bessel functions of first
kind
$I_{0,1}(ae/\lambda)$ would appear \citep{Bur88}.
Let us assume $\lambda \gtrsim ae$;
 with this choice, we can safely use
\eqi\exp\left(\rp{ae}{\lambda}\cos E\right)\approx
1+\rp{ae}{\lambda}\cos E.\lb{approx}\eqf In the small eccentricity
approximation we, thus, get \eqi
\dot\varpi\approx\rp{\alpha\sqrt{GMa}}{2\lambda^2}\exp\left(-\rp{a}{\lambda}\right),\lb{res}\eqf
up to terms of order $\mathcal{O}(e^2)$. Expressions analogous to
\rfr{res} can be found in \citep{Bur88,Tal88,Rey05}, in which quantities proportional to the perihelion
advance after one orbital revolution were worked out\footnote{They are
$2\pi\dot\varpi/n$ \citep{Bur88,Tal88} and $\dot\varpi/n$ \citep{Rey05}. }, and in \citep{Ser06b} where the the perihelion secular rate was calculated up to $O(e^4)$.

\section{Constraining the range and the strength of a Yukawa-like fifth force with planetary perihelia}
\begin{figure}
\begin{center}
\includegraphics[width=13cm,height=11cm]{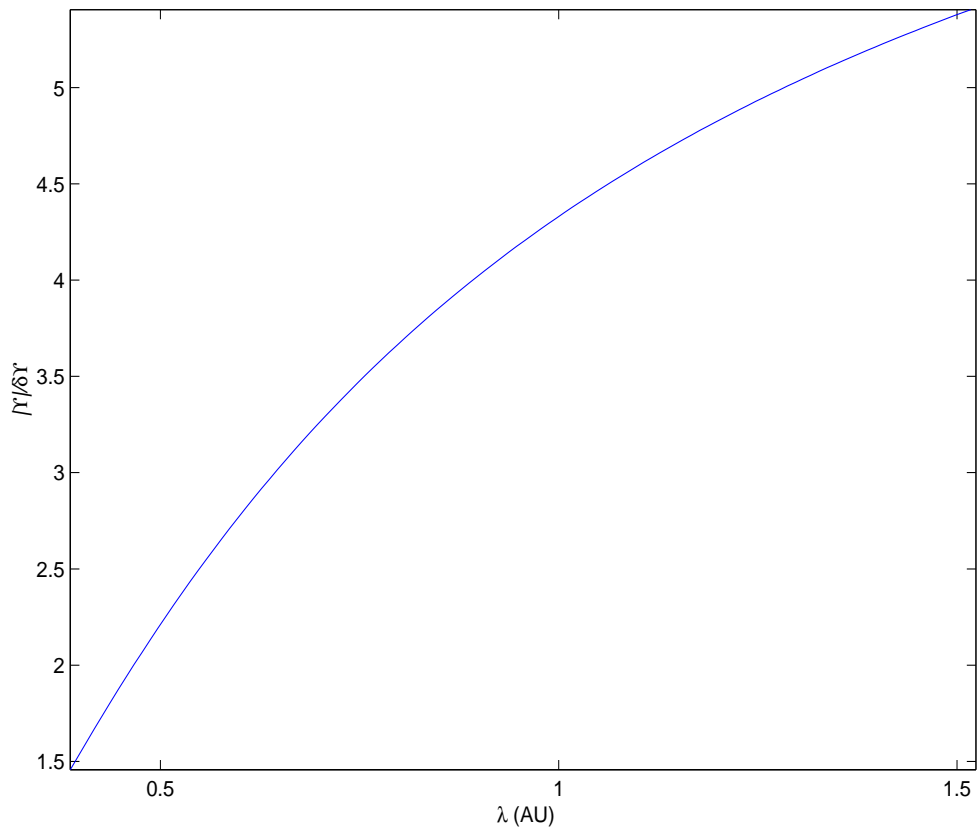}
\end{center}
\caption{\label{figura1} $|\Upsilon|/\delta\Upsilon$ from the data of the Earth and Mercury over a range $a_{\rm Mer}< \lambda < \ a_{\rm Mar}$. Values of $\lambda>a_{\rm Mar}$ are ruled out at an even larger number of $\sigma$.}
\end{figure}

The formula of \rfr{res} is very useful because it allows us to
get important information on the size of $\lambda$. Indeed, let us
write down \rfr{res} for a pair of planets, say A and B, and take
their ratio: by assuming that both $\alpha$ and $\lambda$ do not vary with distance we get
\eqi  \rp{\dot\varpi^{(\rm A)}}{\dot\varpi^{(\rm B)}}=\sqrt{\rp{  a_{\rm A}  }{ a_{\rm B}  }}\exp\left(\rp{a_{\rm B} - a_{\rm A}}{\lambda}\right).\lb{rat}\eqf
Note that the ratio of the rates of perihelia due to a Yuakawa-like interaction is independent of $\alpha$ not only when the approximated expression of \rfr{res} is used, but also when the general expression with the Bessel function  \citep{Bur88} is adopted.
By defining \eqi \Pi\equiv \rp{\dot\varpi^{(\rm A)}}{\dot\varpi^{(\rm B)}},\eqf
and \eqi\Theta(\lambda)\equiv \sqrt{\rp{  a_{\rm A}  }{ a_{\rm B}  }}\exp\left(\rp{a_{\rm B} - a_{\rm A}}{\lambda}\right)\eqf it is possible to construct
\eqi\Upsilon(\lambda)\equiv \Pi-\Theta(\lambda);\eqf if, for a given range of values  of $\lambda\gtrsim ae|_{\rm A/B}$
$|\Upsilon|$ turns out to be incompatible with zero, within the errors, that range for $\lambda$ must be discarded. Note that our analysis is independent  of $\alpha$, assumed to be nonzero, of course.
The uncertainty in $\Upsilon$ can be conservatively assessed as
\eqi\delta \Upsilon(\lambda)\leq\delta\Pi + \delta\Theta(\lambda),\eqf with
\eqi\delta\Pi\leq \left|\Pi\right|\left[ \rp{\delta\dot\varpi^{(\rm A)}}{|\dot\varpi^{(\rm A)}|} +
\rp{\delta\dot\varpi^{(\rm B)}}{|\dot\varpi^{(\rm B)}|} \right],\lb{errperi}\eqf
\eqi\delta\Theta(\lambda)\leq \Theta(\lambda)
\left(\left|\rp{1}{2a_{\rm A}}-\rp{1}{\lambda}\right|\delta a_{\rm A}+
\left|-\rp{1}{2a_{\rm B}}+\rp{1}{\lambda}\right|\delta a_{\rm B}\right).\eqf
The linear sum of the individual errors in \rfr{errperi} accounts for the existing correlations among the estimated perihelia corrections, which reach a maximum of about 20$\%$ for Mercury and the Earth (Pitjeva, private communication 2005).
For A=Earth and B=Mercury, Table \ref{tavola}
\begin{table}
\caption{Estimated semimajor axes $a$, in AU (1 AU$=1.49597870691\times 10^{11}$ m) \citep{Pit05b}, and phenomenologically estimated corrections to the Newtonian-Einsteinian perihelion rates, in arcseconds per century (\asec), of Mercury, the Earth and Mars \citep{Pit05a}. Also the associated errors are quoted: they are in m for $a$ \citep{Pit05b} and in \asec\ for $\dot\varpi$ \citep{Pit05a}. For the semimajor axes they are the formal, statistical ones, while for the perihelia they are realistic in the sense that they
were obtained from comparison of many different
solutions with different sets of parameters and observations (Pitjeva, private communication 2005). The results presented in the text do not change if $\delta a$ are re-scaled by a factor 10 in order to get more realistic uncertainties.}\label{tavola}

\begin{tabular}{ccccc} \noalign{\hrule height 1.5pt}
Planet & $a$ (AU) & $\delta a$ (m)  & $\dot\varpi$ (\asec) & $\delta\dot\varpi$ (\asec) \\
\hline
Mercury & 0.38709893 & 0.105 & -0.0036 & 0.0050\\
Earth & 1.00000011 & 0.146 & -0.0002 & 0.0004 \\
Mars & 1.52366231 & 0.657 & 0.0001 & 0.0005\\

\hline

\noalign{\hrule height 1.5pt}
\end{tabular}

\end{table}
and Figure \ref{figura1}
tell us that $\lambda \approx a_{\rm Mer}$  is not allowed at about $1.5-\sigma$ level; for larger heliocentric distances the constraints are quite tighter, exceeding the $3-\sigma$ level. Although the inspected range for $\lambda$ ends at Mars in Figure \ref{figura1}, it turns out that larger values, far beyond the Solar System boundaries, are ruled out as well at about $8-\sigma$ level. Our results are unaffected by re-scaling by a factor 10 the formal errors in the semimajor axes.

After having discovered that $\lambda$ cannot exceed the semimajor axis of Mercury, let us  now further constrain it. From \rfr{rat} it can be obtained
\eqi\lambda = \rp{a_{\rm B}-a_{\rm A}}{\ln{\left(\sqrt{\rp{a_{\rm B}}{a_{\rm A}}}\Pi\right)}}\lb{lam}.\eqf
It turns out that the major sources of error are the estimated extra-rates of perihelia through their ratio $\Pi$, so that
\eqi \delta\lambda\leq\left|\rp{a_{\rm B}-a_{\rm A}}{\Pi\ln^2{\left(\sqrt{\rp{a_{\rm B}}{a_{\rm A}}}\Pi\right)}}\right|\delta\Pi.\eqf
For A=Earth, B=Mercury we have
\eqi \lambda = 0.182\pm 0.183\ {\rm AU},\lb{lambda}\eqf
which is marginally compatible with zero. Note that the use of \rfr{res}, from which \rfr{rat} and \rfr{lam} come, can be a posteriori justified because for the Earth and Mercury the obtained value for $\lambda$ yields $ae/\lambda <1$.
%

The result of \rfr{lambda} allows us to constrain $\alpha$ as well. Indeed, in the case of Mars we have
\eqi\rp{ae}{\lambda}=0.78,\eqf so that the approximation of \rfr{approx}, and the formula of \rfr{res} based on it, hold. Thus, from \rfr{lambda} and the values of Table \ref{tavola} for Mars we get
\eqi \alpha=\rp{2\lambda^2\dot\varpi}{\sqrt{GMa}}\exp{\left(\rp{a}{\lambda}\right)}=2\times 10^{-10}.\lb{alfa}\eqf
The uncertainty can be evaluated as
\eqi \delta\alpha\leq\alpha\left(\rp{1}{\lambda}\left|2-\rp{a}{\lambda}\right|\delta\lambda+\rp{1}{|\dot\varpi|}\delta\dot\varpi\right)=1.3\times
10^{-9}.\eqf
Also in this case, $\alpha$ is compatible with zero.
Such constrains on $\alpha$ are less tight than those obtained, e.g., in \citep{Ior07a,Ser06b}, but the authors of such works made use of values of $\lambda$ which the present analysis has ruled out.

\section{Conclusions}
In this paper we put on the test the hypothesis that modifications of the Newtonian inverse-square law, parameterized  in terms of a Yukawa-like correction, can occur over astronomical scales by using the corrections to the Newtonian-Einsteinian secular rates of the perihelia of Mercury and the Earth phenomenologically estimated, in the least-square sense, with the EPM2004 ephemerides by \citet{Pit05a}. 

By taking their ratio  we found that the range parameter $\lambda$ of a Yukawa-like fifth force cannot exceed about 0.18 AU.
The determined extra-precession of the perihelion of Mars yielded an upper bound on $\alpha$ of $10^{-9}$, which is compatible with other estimates obtained with other approaches. The values obtained for both $\lambda$ and $\alpha$ are compatible with zero; moreover, the results presented here are left unaffected  by re-scaling the uncertainties in the estimated Keplerian orbital elements by a factor 10 in order to evaluate them more realistically. 

If and when corrections to the secular rates of perihelia will be estimated by other teams of astronomers,  more complete and extensive tests could be performed.

Another approach which could be followed consists in introducing an ad-hoc Yukawa-type term in the dynamical force models of the ephemerides-generating routines and repeating the global fit of the whole Solar System data set by estimating, among other things, also the parameters in terms of which the new force is expressed.


\end{document}